\documentclass[epj]{svjour}
\usepackage{graphicx}
\usepackage{color}
\usepackage{amssymb}
\usepackage{amsmath}
\newcommand{\be}{\begin{equation}}
\newcommand{\ee}{\end{equation}}
\newcommand{\bea}{\begin{eqnarray}}
\newcommand{\eea}{\end{eqnarray}}
\newcommand{\nn}{\nonumber}

\newcommand{\noi}{\noindent}

\begin{document}
\title{Radial Flow in Non-Extensive
Thermodynamics and Study of Particle Spectra at LHC
in the Limit of Small $(q-1)$}
\author{Trambak Bhattacharyya\inst{1} \and Jean Cleymans\inst{2} \and Arvind Khuntia\inst{1} \and Pooja Pareek\inst{1} \and Raghunath Sahoo\inst{1}\thanks{Corresponding Author E-mail: Raghunath.Sahoo@cern.ch}}

\vspace{0.2in}

\institute{Discipline of Physics, School of Basic Sciences, Indian Institute of
Technology Indore, M.P. 452020, India \and
UCT-CERN Research Centre and Department of Physics, University of Cape Town, Rondebosch 7701, South Africa }

\date{\today}

\abstract{
We expand the Tsallis distribution in a Taylor series of powers of ($q-1$), where $q$ is the Tsallis
parameter, assuming $q$ is very close to 1. This helps in
  studying the degree of deviation of transverse momentum spectra and
  other thermodynamic quantities from a thermalized Boltzmann distribution.
After checking thermodynamic consistency, we provide analytical results for the Tsallis distribution in 
the presence of collective flow up to the first order of ($q-1$). The formulae are compared with the experimental data.
}
\authorrunning{T. Bhattacharyya {\it et al.}}
\titlerunning{Non-Extensive Thermodynamics}
\maketitle


\section{Introduction}
\label{intro}

It is now a standard  practice to use the Tsallis distribution~\cite{tsallis} for  
describing the transverse momentum distributions at high energies. This has been pioneered
by the PHENIX and  STAR collaborations~\cite{STAR,PHENIX1,PHENIX2} at the 
Relativistic Heavy Ion Collider (RHIC) at BNL and by the 
ALICE, ATLAS and CMS collaborations~\cite{ALICE_charged,ALICE_piplus,CMS1,CMS2,ATLAS,ALICE_PbPb} at the 
Large Hadron Collider (LHC) at CERN. The Tsallis distribution is
successful in explaining the experimental transverse momentum distribution,
longitudinal momentum fraction distribution as well as rapidity distribution of
hadrons off the $e^+e^-$ as well as $p-p$ collisions
\cite{e+e-,R1,R2,R3,ijmpa,plbwilk,marques}.
The form of the Tsallis distribution used in this paper has been described in detail 
previously~\cite{worku1,worku2,worku3,azmi1} and has the advantage of being 
thermodynamically consistent. 
There is clear evidence for a mild energy dependence of the
parameters $q$ and $T$~\cite{worku3}. Also, initially there were indications
that the values obtained for the parameters $q$ and $T$
were consistent with each other for different particle species~\cite{marques,worku2}.
Different  conclusions have been reached in the literature~\cite{TBW,chin,biro12009,biro22009}, albeit
using slightly different formalisms and approaches,  and
a  more detailed analysis is still outstanding to prove this
beyond doubt.

\section{Review of the Main Ingredients of the  Model}
For completeness we recall here the main ingredients.

\indent The relevant thermodynamic quantities can be written as integrals over the following distribution function: 
\begin{equation}
f = \left[1 + (q-1)\frac{E-\mu}{T}\right]^{-\frac{1}{q-1}}  .
\label{tsallis}
\end{equation}
It can be shown \cite{worku2} that the 
entropy, $S$,  particle number, $N$, energy density, $\epsilon$, and the pressure, $P$.\\
are given by,
\begin{eqnarray}
S &=& - gV\int\frac{d^3p}{(2\pi)^3}\left[f^{q}{\rm ln}_{q}f - f\right],\label{entropy}\\
N &=& gV\int\frac{d^3p}{(2\pi)^3} f^{q},\label{Number}\\
\epsilon &=& g\int\frac{d^3p}{(2\pi)^3}~E~f^{q},\label{epsilon}\\
P &=& g\int\frac{d^3p}{(2\pi)^3}\frac{p^{2}}{3E}~f^{q}.\label{pressure}
\end{eqnarray}
where $V$ is the volume and  $g$ is the degeneracy factor. \\
\indent The function appearing in Eq.~\eqref{entropy} 
is often referred to as q-logarithm and is defined by
$$
\ln_{q}(x)\equiv\frac{x^{1-q} - 1}{1-q}.
$$

\indent The first and second laws of
thermodynamics lead to the following two 
differential relations:
\begin{eqnarray}
d\epsilon=T~ds +\mu~dn,\label{eq1}\\
dP=s~dT + n~d\mu.\label{eq2}
\end{eqnarray}
where, $s = S/V$ and $n = N/V$ are the entropy and particle number 
densities, respectively.\\
\indent It is seen that if we use $f^q$ in stead of $f$ to define the
thermodynamic variables, the above equations satisfy the
thermodynamic consistency conditions which require that
the following relations to be satisfied:
\begin{eqnarray}
T=\left.\frac{\partial\epsilon}{\partial s}\right|_n,\label{temp}\\
\mu=\left.\frac{\partial\epsilon}{\partial n}\right|_s,\label{mu}\\
n=\left.\frac{\partial P}{\partial \mu}\right|_T,\label{number}\\
s=\left.\frac{\partial P}{\partial T}\right|_\mu.\label{ent}
\end{eqnarray}
\indent Eq.~\eqref{temp}, in particular, shows that the variable $T$ appearing in Eq.~\eqref{tsallis} can indeed
be identified as a thermodynamic temperature and is more than just another parameter. 
It is straightforward to show that these relations are indeed satisfied~\cite{worku2}.

Based on the above expressions the particle distribution can be rewritten, using variables appropriate for 
high-energy physics as 

\bea
\frac{dN}{dp_T dy} &=& \frac{gV}{(2\pi)^2} p_T m_T \mathrm{cosh}y \nonumber\\
                   &&  \left(1+(q-1)\frac{m_T \mathrm{cosh}y
-\mu}{T}\right)^{-\frac{q}{q-1}}
\label{tsallisptdist}
\eea
It can be shown that at central rapidity,
 $y=0$, one can obtain the transverse
momentum distribution
in terms of the central rapidity density, $dN/dy|_{y=0}$, as
the volume dependence can be replaced by a dependence on $dN/dy|_{y=0}$
using

\begin{eqnarray}
\left.\frac{dN}{dy}\right|_{y=0}
&=& \frac{gV}{(2\pi)^2} \left[ 1 + (q-1)\frac{m-\mu}{T}\right]^{-\frac{1}{q-1}}\nonumber\\
&& \frac{T^3}{(2q-3)(q-2)}\nonumber\\
&& \left[ 2 - (q-2)\left(\frac{m-\mu}{T}\right)^2 + 2\frac{m-\mu}{T}\right.\nonumber\\
&& -2\frac{\mu}{T}(2q-3)\left(1 + \frac{m-\mu}{T}\right) \nonumber\\
&& \left.+\frac{\mu^2}{T^2}(2q-3)(q-2)\right]  .
\label{dndy}
\end{eqnarray}

This leads to the following expression and  generalizes the expression given in \cite{Poland} to 
non-zero values of the chemical potential $\mu$ (see Appendix \ref{eq1314} for
an outline of the derivation of Eq. \eqref{fittingformula}):

\begin{eqnarray}
&& \frac{dN}{dp_Tdy}\bigg|_{y=0} = \frac{p_T\,m_T}{T} \frac{dN}{dy}\bigg|_{y=0}
\bigg[1+(q-1)\frac{m_T-\mu}{T} \bigg]^{-\frac{q}{q-1}}\nonumber\\
&\times&\frac{(2-q)\delta}{(2-q)d^2 + 2dT + 2T^2 +2\mu\delta(T+d)+\mu^2\delta (2-q)}
\nonumber \\
&\times&\bigg[1+(q-1)\frac{d}{T} \bigg]^{\frac{1}{q-1}}.
\label{fittingformula}
\end{eqnarray}

\noindent where the abbreviations $d\equiv m-\mu$ and $\delta \equiv 3-2q$ have been used.

In all fits to transverse momentum spectra,  the parameter $q$ turns out to be
very close to 1~\cite{worku3,azmi1}. In fact, the value of the non-extensive
parameter $q$ for high energy collisions is found to be $1\leq q\leq 1.2$
\cite{e+e-,beck}. In the limit where $q$ is exactly 1, Eq.~\eqref{tsallisptdist}
reduces to the standard exponential function appearing in the Boltzmann
distribution. It is  therefore  useful  to expand the above expressions in a
Taylor series in ($q-1$) and  see how the deviations from a Boltzmann
distribution develop.  Such an expansion has been considered previously
in~\cite{osada-wilk,alberico}. The present paper develops a more systematic
analysis than the previous ones and considers a slightly different form of the
Tsallis distribution, having an extra power of $q$, because it is consistent
with basic thermodynamic relations.

The aim of this paper is to develop a Taylor expansion of  Eq.
\eqref{tsallisptdist} in  ($q-1$) based on the fact that
($q-1$) $<<1$ (see for example \cite{JCConf}). The conditions of validity
of such an expansion for pure Tsallis distribution (Eq. \eqref{tsallis}) is
$|1-q|E/T<1$. Apart from this, up to first order in $(q-1)$ an additional
condition $|1-q|(E/T)^2<2$ must be satisfied \cite{osada-wilk}. The condition
of validity for expansion up to order $(q-1)^2$ term will be $|1-q|^2(E/T)^3<3$.
The expansion to higher orders has also been considered in~\cite{tsallis-iran} 
in the framework of an analysis of  quasi-additivity for the Tsallis entropy
for different subsystems.

 The Taylor
expansion is useful
as a mathematical tool because it breaks the Tsallis distribution in a series of
($q-1$) containing powers of energy $E$.
 Now, the advantage we get is, it will
be easier to consistently include the effect of flow on the Tsallis distribution
just by making a substitution $E\rightarrow p^\mu u_{\mu}$, for a collective four-velocity 
$u^{\mu}$ of particles with representative four-momentum $p^{\mu}$ \cite{PShukla,BDe1}.

There have been earlier attempts to include the effect of collectivity in the dynamics
of the particles following the Tsallis distribution in the form of Tsallis-Blast
Wave (TBW) formalism \cite{TBW}, or in Refs. \cite{PShukla,urmossy}.
In all the cases, an ansatz of fluid four-velocity is taken and energy is
replaced by the scalar product $p^{\mu}u_{\mu}$. The inclusion of flow
inside non-extensive statistics reduces the value of $q$ \cite{dhananjay} since
some degree of non-extensivity is shared by the dynamics. Also, whenever we
have an inhomogeneous thermodynamic system, with regions having different
temperatures and exchanging heat with the bigger system, we can define an
effective temperature $T_{\mathrm{eff}}$ which is affected by energy transfer
only when $q\ne1$ \cite{wilkteff,epjateff,jpgteff}. The variation of effective
temperature with $q$ is seen in \cite{ijmpeteff}. Another important observed
phenomenon like
$m_T$ scaling is affected by flow. In $p-p$ or $p-A$ collisions, the particle
spectra for different hadrons are having the same slope parameter $T$ and this
phenomenon is known as $m_T$ scaling \cite{mtscaling}. Due to this scaling
behaviour, the particle species cannot be identified just by looking at the
spectrum. This behaviour will be manifested in particle distributions following
Eq. \eqref{tsallisptdist}. But, because of the inclusion of collectivity the
particle species start having different slope parameters. The average collective
velocity as well as the mass of particles will contribute to the slope parameter
and hence the scaling is broken \cite {hicbook}.
It is possible that  $q$  represents the  joint action of many  dynamical
factors~\cite{tsallis-japan}
and it cannot be excluded at present that when these are accounted for, the
factor $q-1$ could well become zero.
The values of the parameters are  clearly sensitive to the details of the
hadronization
mechanism~\cite{biro2015,biro-entropy}.


The TBW formalism considers an implicit dependence of fluid rapidity on the
velocity; but in the present case we consider a cylindrical geometry and
velocity comes explicitly in the calculations. In contrast to the numerical
treatment in \cite{TBW}, we provide an analytical formula for the Tsallis
distribution expanded up to $\mathcal{O}(q-1)$ in presence of flow. In the next
section, we derive the Taylor expansion for the Tsallis distribution in the
series of ($q-1$) and we compare the order by order deviation from a thermalized
Boltzmann distribution of transverse momentum spectra in hadronic and nuclear
collisions. In Sec. 4 we find out the expressions for number density ($n$),
pressure ($P$) and energy density ($\epsilon$) for a system with non-zero
chemical potential ($\mu$). We verify the thermodynamic relationship $n=\partial
P/\partial \mu$. In Sec. 5 we derive an analytical formula including flow up to
$\mathcal{O}(q-1)$ of a Taylor series expansion of the Tsallis distribution and
show its application to experimental data. Finally we summarize the paper in
Sec. 6.

\begin{figure}[h]
\includegraphics[width=0.9\columnwidth, height = 14.0cm]{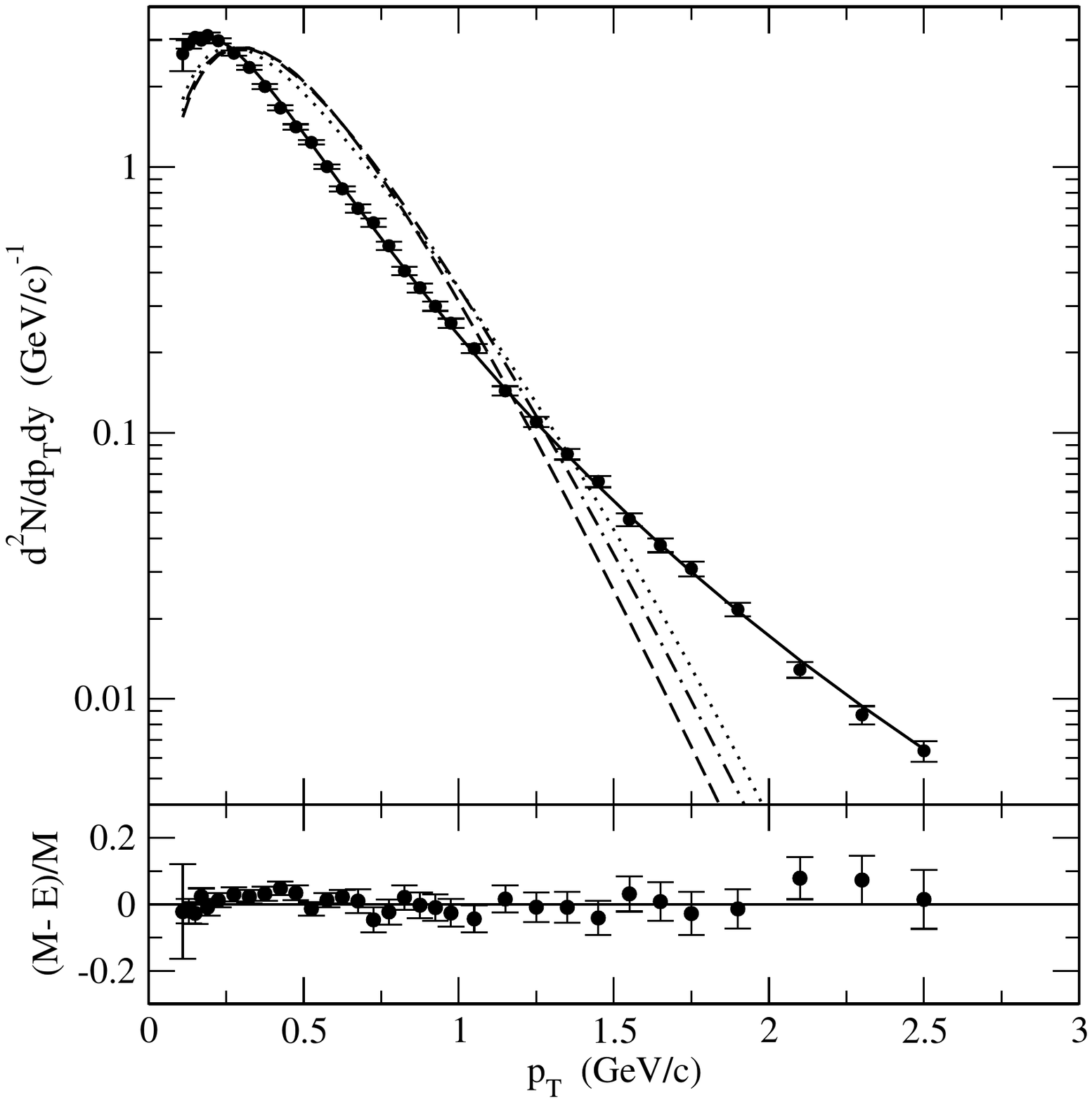}
\caption{
Fits to the normalized differential yields of $\pi^+$ as measured by the 
ALICE collaboration in  $p-p$ collisions at $\sqrt{s}$ = 0.9 TeV~\cite{ALICE_piplus} 
fitted with the Tsallis (solid line) 
and Boltzmann distributions (dashed line). Also shown are fits with the Tsallis distribution keeping terms to first
(dash-dotted line) and second order in $(q-1)$ (dotted line). 
The lower part of the figure shows the difference between model (M) and experiment (E) normalized to the model (M) values.
}
\label{fig:pp}
\end{figure}
  
\begin{figure}[h]
\includegraphics[width=0.9\columnwidth, height = 8.0cm]{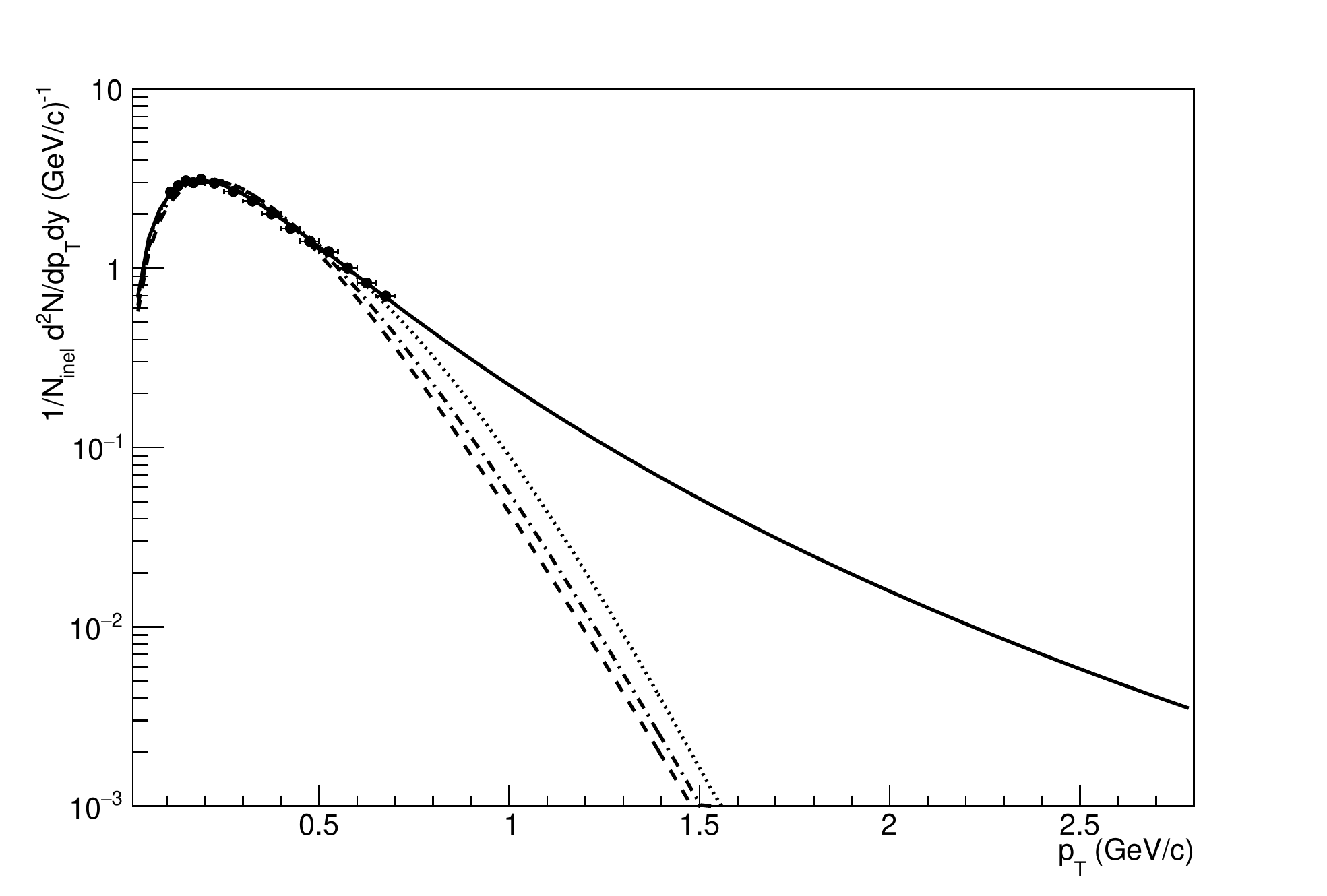}
\caption{
Fits to the normalized differential yields of $\pi^+$, including only results at small values of the transverse momentum, as measured by the 
ALICE collaboration in  $p-p$ collisions at $\sqrt{s}$ = 0.9 TeV~\cite{ALICE_piplus} 
fitted with the Tsallis (solid line) 
and Boltzmann distributions (dashed line). Also shown are fits with the Tsallis distribution keeping terms to first
(dash-dotted line) and second order in $(q-1)$ (dotted line). 
}
\label{fig:pp_smallpt}
\end{figure}

\section{Momentum Distributions to First and Second Order in $(q-1)$}
Assuming the parameter $q$ to be close to 1, as is the value for many cases in
high energy physics, the modified Tsallis
distribution, $f^q$, appearing in the expressions for the thermodynamic
quantities can be expanded in a Taylor series in $q-1$ with the following result
(see appendix 
\ref{apptaylorsexp} for a detailed
derivation):
\begin{eqnarray}
&&\left[1+(q-1)\frac{E-\mu}{T}\right]^{-\frac{q}{q-1}}\nonumber\\
&\simeq& \mathrm{e}^{-\frac{E-\mu}{T}}\left\{ 1 +  
(q-1)\frac{1}{2}\frac{E-\mu}{T}\left( -2 + \frac{E-\mu}{T}\right)\right.\nonumber\\
&+&\frac{(q-1)^2}{2!}\frac{1}{12}\left[\frac{E-\mu}{T}\right]^2
\left[ 24 - 20\frac{E-\mu}{T} +3\left( \frac{E-\mu}{T}\right)^2\right]\nonumber\\
&+& \mathcal{O}\left\{(q-1)^3\right\}\nonumber\\
&+&\left. ... \right\}
\label{taylor}
\end{eqnarray}

%
\begin{figure}[ht]
\includegraphics[width=\columnwidth, height = 14.0cm]{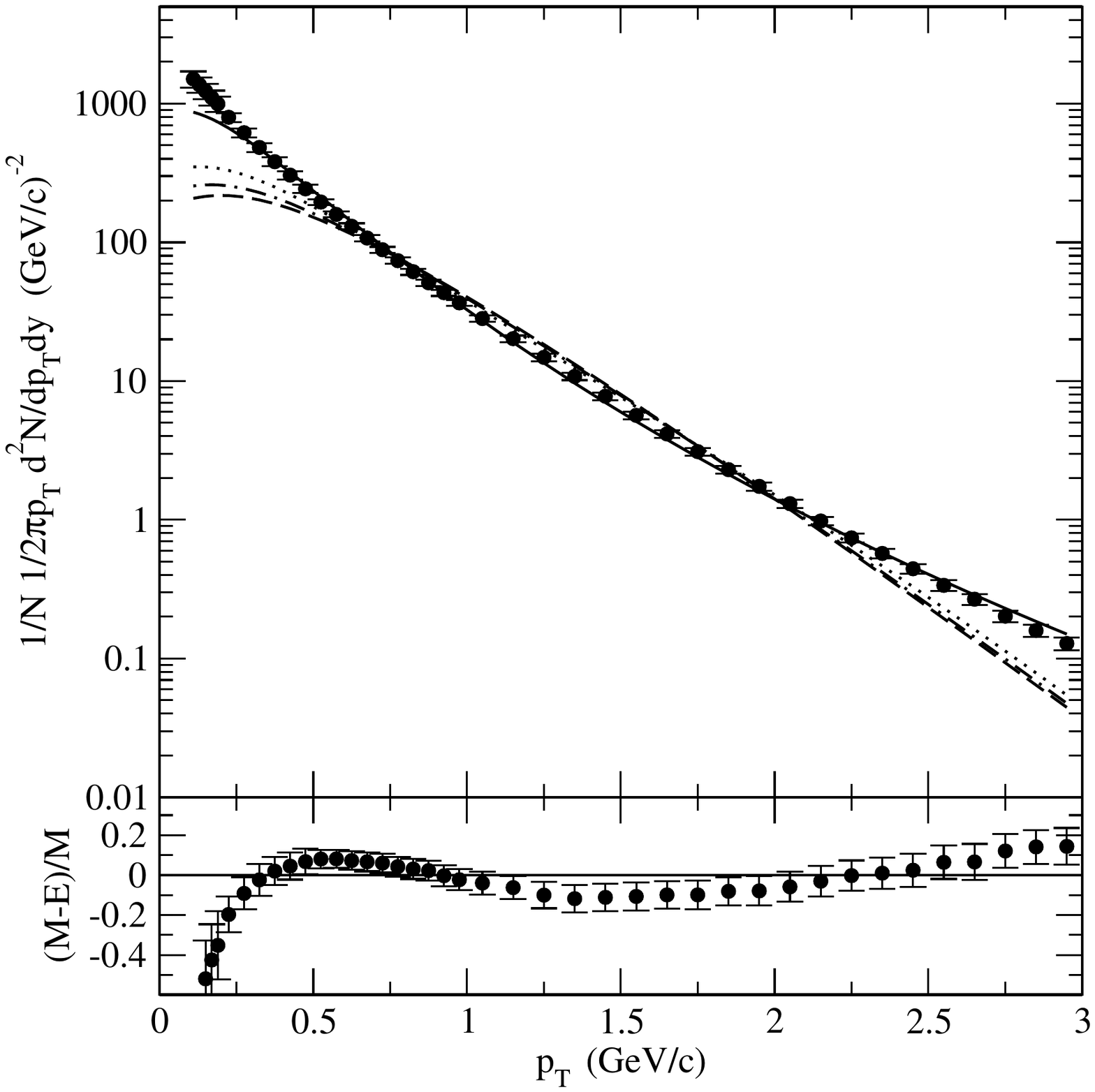}		
\caption{\label{ALICEPbPb} Fits to the normalized differential $\pi^-$ yields as measured by the 
ALICE collaboration in  $(0-5)\%$ Pb-Pb collisions at $\sqrt{s_{\rm
  NN}}$ = 2.76 TeV~\cite{ALICE_PbPb} fitted with the Tsallis 
(solid line) and Boltzmann distributions (dashed line). Also shown are fits with the Tsallis distribution keeping terms 
to first (dash-dotted line) and second order in $(q-1)$ (dotted line).
The lower part of the figure shows the difference between model (M) and experiment (E) normalized to the model (M) values.
}
\label{fig:pbpb}
\end{figure}

This result can be used for the invariant yield of particles 
if it is written in terms of  the Tsallis distribution,

\begin{eqnarray}
E\frac{dN}{d^3p}=\mathcal{C}E\left[1+(q-1)\frac{E-\mu}{T}\right]^{-\frac{q}{q-1}}
\end{eqnarray}
where 
$\mathcal{C} \equiv gV/(2\pi)^3$.
 Let us use the following notations: 
($q-1$)$\equiv x$; $(E-\mu)/T=\Phi$ and $1+(q-1)\frac{E-\mu}{T} =
1+x~\Phi=f(x)$.
Hence, the expansion of the Tsallis distribution up to $\mathcal{O}(x^2)$ can
be written as:

\bea
E\frac{dN}{d^3p} &\simeq& \mathcal{C}E   e^{-\Phi} 
+ \mathcal{C}E \frac{x}{1!}  \frac{\Phi}{2} 
\left(-2+\Phi \right) e^{-\Phi} \nonumber\\
&+& \mathcal{C}E \frac{x^2}{2!} \frac{\Phi^2}{12} \left(24-20\Phi
+3 \Phi^2 \right)  e^{-\Phi} 
\eea
Hence one obtains
\begin{eqnarray}
\frac{dN}{p_Tdp_Tdyd\phi} &\simeq& \mathcal{C}E   e^{-\Phi} 
+ \mathcal{C}E \frac{x}{1!}  \frac{\Phi}{2} 
\left(-2+\Phi \right) e^{-\Phi} \nonumber\\
&+& \mathcal{C}E \frac{x^2}{2!} \frac{\Phi^2}{12} \left(24-20\Phi
+3 \Phi^2 \right)  e^{-\Phi} 
\label{order12}
\end{eqnarray}

Since, we use a modified form of Tsallis distribution, a comparison with
similar work will be worthwhile at this point. With this aim, we 
integrate over the rapidity variable to compare the transverse mass spectrum
obtained from the present approach with that obtained in Ref. \cite{alberico}:
\begin{eqnarray}
\frac{dN}{m_Tdm_T} &=& \frac{gV}{2\pi^2} m_T \left[ K_1 \left(\frac{m_T}{T}\right) \mathrm{e}^{\frac{\mu}{T}} \right. \nn\\
&&- \frac{q-1}{2} \frac{m_T}{T} \left\{    K_0 \left(\frac{m_T}{T}\right) + K_2 \left(\frac{m_T}{T}\right) \right\} \nn\\
&+&  \frac{q-1}{8} \left(\frac{m_T}{T}\right)^2 \left\{  K_3 \left(\frac{m_T}{T}\right) + 
3 K_1\left(\frac{m_T}{T}\right) \right\} \nn\\
&& + \frac{\mu (q-1)}{T} K_1 \left(\frac{m_T}{T}\right) + \frac{\mu^2 (q-1)}{2T^2} K_1 \left(\frac{m_T}{T}\right) \nn\\
&& - \left. \frac{\mu m_T (q-1)}{2T^2}  \left\{    K_0 \left(\frac{m_T}{T}\right) + K_2 \left(\frac{m_T}{T}\right) \right\} \right] \nn\\
&&+\mathcal{O}((q-1)^2) +\ldots 
\label{trmassspec}
\end{eqnarray}
where $K_n$s are the modified Bessel's functions of second kind (see Appendix
\ref{app:flow}). In principle, while examining the transverse spectra, the
rapidity integration should be within a maximum value $y_{max}$, say. But,
owing to the presence of the term $e^{-m_T\mathrm{cosh}y/T}$, the
integrand drops down very fast with increasing $y$. And so, according to the
standard practice, we can effectively replace the rapidity
integration from 0 to $y_{max}$ to 0 to $\infty$ so that the integration yields
Bessel functions. The first term in the above expression is the well-known
formula for a thermal source with a Boltzmann distribution:

\be
\frac{dN}{p_Tdp_T}=\frac{gV}{2\pi^2} m_T \mathrm{e}^{\frac{\mu}{T}} K_1
\left(\frac{m_T}{T}\right)
\label{boltzspectra}
\ee
In the limit $\mu=0$, the transverse mass spectrum obtained from
Ref. \cite{alberico} (up to $\mathcal{O}(q-1)$) does not contain the term
involving $K_0$ and $K_2$.

The Boltzmann distribution, the pure Tsallis distribution (Eq.
\eqref{tsallisptdist}) and the expansion of the Tsallis distribution up to the
first and second order of $(q-1)$ (Eq. \eqref{order12}) were used to fit the
transverse momentum distributions obtained by the ALICE collaboration. The
results are shown in Fig.~\ref{fig:pp} for p-p collisions at
$\sqrt{s}=0.9$ TeV and in Fig.~\ref{fig:pbpb} for Pb-Pb collisions at
$\sqrt{s_{\rm NN}}=2.76$ TeV. It is well known that the Tsallis fits give
excellent results for p-p collisions but are not very good for Pb-Pb collisions
as can be seen clearly in Fig. (2) where a large deviation at small $p_T$ is
seen.

It can be seen  from the fits to the $p-p$ distribution  that the successive
terms in $(q-1)$ improve the fits but not in a convincing manner. Clearly, it is
the best to use the full Tsallis distribution and not the series expansion. It
might turn out of course that the series expansion could be of use in  a
different situation where the comparison with a Boltzmann
distribution is more relevant.

The fits in the figures were done using the MINUIT package with the  following
numerical  results. 

In Fig. ~\ref{fig:pp}  we show fits to the transverse momentum distribution of
$\pi^+$ in p-p collisions at 900 GeV. For the plain
Tsallis distribution (solid line) the parameters were obtained as being
$T$ = 70.8 MeV,   $q$ = 1.1474. The volume $V$ was determined as corresponding
to a spherical radius  of 4.81 fm. For the Boltzmann distribution (dashed line)
the parameters were determined as being $T$ = 150.2 MeV, while the radius used
to determine the volume was fixed at a value of 2.65 fm.  For the fit using 
the Boltzmann distribution and the  first order term in ($q-1$)
(dashed-dotted line) the values are $T$ = 138.4 MeV, $q$ = 1.035
while the radius is given by 2.80 fm.  In the last case corresponding to Boltzmann plus first and second orders in ($q-1$) (dotted
line) one has $T$ = 121.2 MeV, $q$ = 1.065 and a radius of 3.09 fm.  As is well-known and evident, the fit using the Tsallis 
distribution is very good.

In Fig. ~\ref{fig:pp_smallpt}  we show fits to the small  transverse momentum region of $\pi^+$ in p-p collisions at 900 GeV.
In this case the fits using the Boltzmann distribution are fairly good. The deviation with the Tsallis distribution becomes very prominent only
 for larger 
transverse momenta. For the plain
Tsallis distribution (solid line) the parameters were obtained as being $T$ = 70.8 MeV,   $q$ = 1.145, similar to the full range of $p_T$.
The volume $V$ was determined as corresponding to a spherical radius  of 4.82 fm.
For the Boltzmann distribution (dashed line) the parameters were determined as being $T$ = 104.9 MeV, i.e. much lower than the full range in $p_T$
while the 
radius used to determine the volume was fixed at a value of 3.61 fm.  For the fit using 
the Boltzmann distribution and the  first order term in ($q-1$) (dashed-dotted line) the values are $T$ = 89.9 MeV, $q$ = 1.07
while the radius is given by 4.03 fm.  In the last case corresponding to Boltzmann plus first and second orders in ($q-1$) (dotted
line) one has $T$ = 77 MeV, $q$ = 1.11 and a radius of 4.54 fm.  

In Fig. ~\ref{fig:pbpb} we show fits to the
normalized differential $\pi^-$ yields 
in $(0-5)\%$ Pb-Pb collisions  at $\sqrt{s_{\rm NN}} $ = 2.76 TeV as measured by the ALICE collaboration~\cite{ALICE_PbPb} 
with the Tsallis (solid line) 
and Boltzmann distributions (dashed line). Also shown are fits with the Tsallis distribution keeping terms to first order
(dash-dotted line) and second order in $(q-1)$ (dotted line). 
The lower part of the figure shows the difference between the Tsallis distribution (M) and experiment (E).
It is clear that the best fit is achieved with the full Tsallis distribution, whereas, using the Boltzmann distribution the description is not good. Successive corrections in $(q-1)$ improve the description.
There is a clear deviation at very  low transverse momentum (below 0.5 GeV) and also at higher values above 2.75 GeV.

\section{Thermodynamic Quantities to First Order in $(q-1)$}
The particle density  in Tsallis thermodynamics is given to first order in $(q-1)$ by the following expression:
\begin{equation}
n^B + (q-1)n^1
\end{equation}
where $n^B$ is the standard Boltzmann result for the particle density:
\bea
n^B=\frac{g}{2\pi^2}\mathrm{e}^{\frac{\mu}{T}}T^3 a^2 K_2(a),
\eea
with $a\equiv m/T$,
and the first order expression  in $q-1$ is given by 
\begin{eqnarray}
n^1&=&\frac{g\mathrm{e}^{\frac{\mu}{T}}T^3}{4\pi^2}\left[-6a^2K_2(a)-2a^3 K_1(a)\right.\nonumber\\
&& -4a^2bK_2(a)+3a^3K_3(a)+a^4K_2(a)+a^2 b^2 K_2(a)\nonumber\\
&&\left.-2a^3bK_1(a)\right]  .
\label{density}
\end{eqnarray}

\begin{figure}[ht]
\includegraphics[width=\columnwidth, height = 7.0cm]{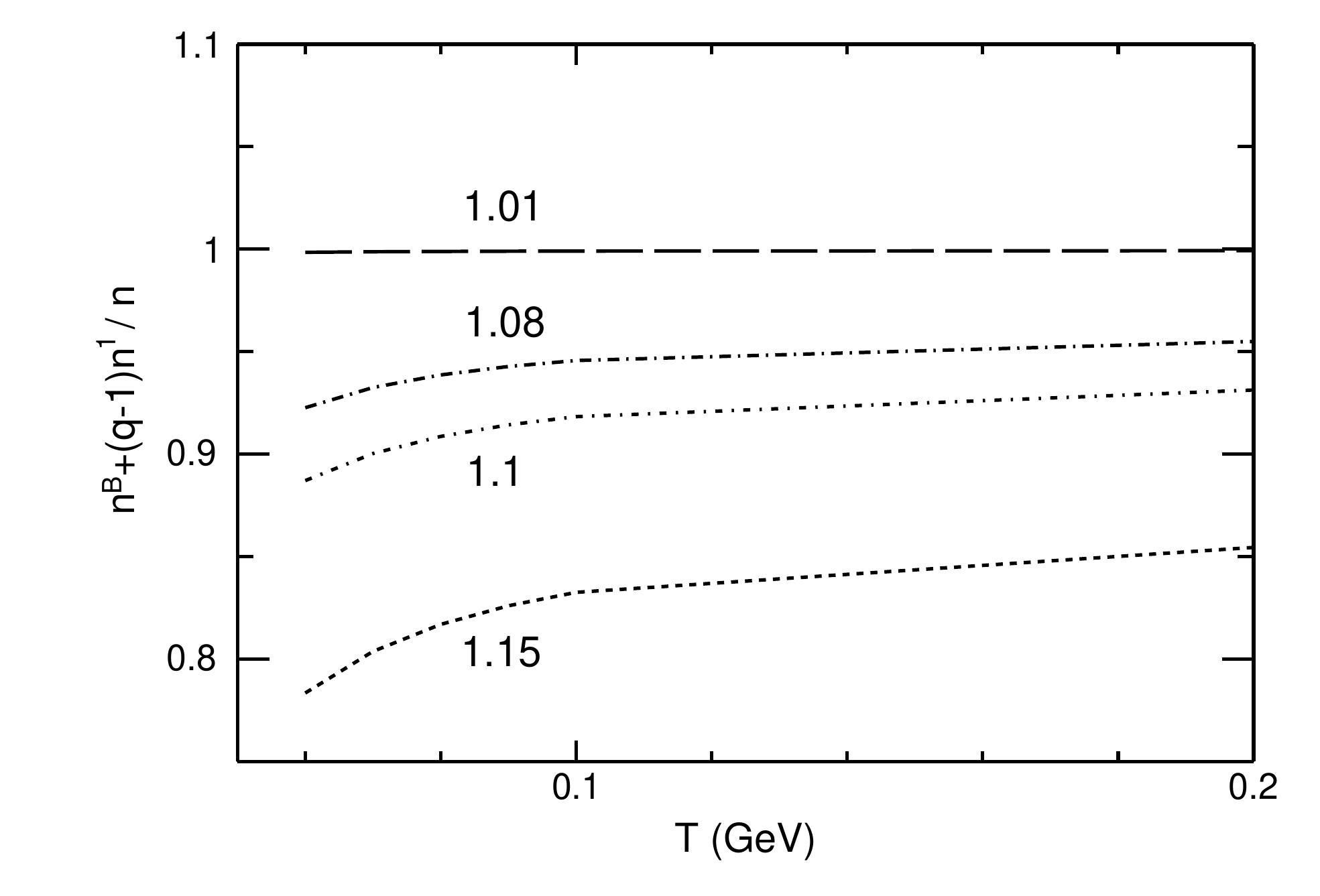}			
\caption{\label{particle_density_tsallis_caption} 
The ratio of the particle density calculated to first order in $(q-1)$
normalized to the  particle density of a Tsallis gas as
a function of the temperature for different values of the parameter $q$.  The
mass is taken as being the pion mass. The values of the parameter $q$ are 1.01
for the dashed line, 1.08 for the dot-(long)dashed line, 1.1 for the
dot-(short)dashed line and 1.15 for the dotted line.}
\label{fig:particle_density_tsallis}
\end{figure}

In Fig.~\ref{fig:particle_density_tsallis} we show the ratio of the particle
density to first order in $(q-1)$ to the full particle density as given by the
Tsallis distribution, $(n^B+(q-1)n^1)/n$ for several values of $q$  indicated in
the figure as a function of the temperature $T$.\\
It can be seen that the expansion in $(q-1)$ is excellent if $(q-1) = 0.01$
but rapidly deviates from the full Tsallis distribution for larger values of
$q$. Already for $(q-1) \approx 0.1$ the deviations are of the order of 10 \% as
can be seen from Fig.~\ref{fig:particle_density_tsallis}.

\begin{figure}[ht]
\includegraphics[width=\columnwidth, height = 7.0cm]{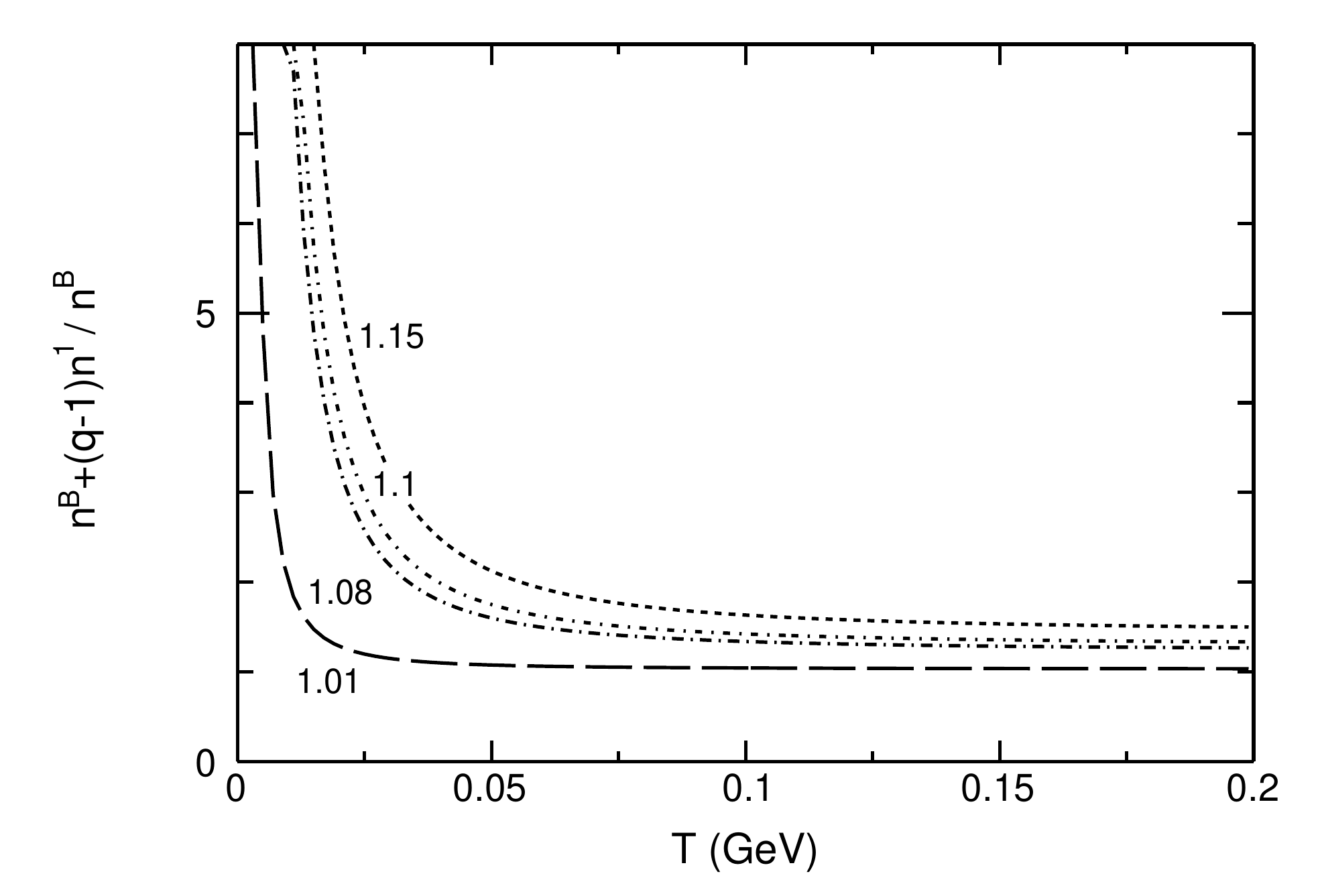}		
\caption{\label{particle_density_boltzmann_caption} 
The ratio of the particle density calculated to first order in $(q-1)$ normalized to the  particle density of a Boltzmann gas as
a function of the temperature for different values of the parameter $q$.  The
mass is taken as being the pion mass. The values of the parameter $q$ are 1.01
for the dashed line, 1.08 for the dot-(long)dashed line, 1.1 for the
dot-(short)dashed line and 1.15 for the dotted line.}
\label{fig:particle_density_boltzmann}
\end{figure}

For comparison we show in Fig. \ref{fig:particle_density_boltzmann}, the first order expansion compared to the
Boltzmann expression,   $(n^B+(q-1)n^1)/n^B$, again as a function of
the temperature $T$ for several values of the parameter $q$. In this case the deviations are most pronounced for small values of the temperature. 
  

Correspondingly, the energy density is obtained as:
\begin{equation}
 \epsilon^B + (q-1)\epsilon^1
\end{equation}

\bea
\epsilon^B =  \frac{ g \mathrm{e}^{\frac{\mu}{T}} T^4}{2\pi^2} (3a^2K_2(a)+a^3K_1(a))
\eea

\bea
\epsilon^1&=&\frac{g\mathrm{e}^{\frac{\mu}{T}}T^4}{4\pi^2}\left[9a^3K_3(a)+4a^4K_2(a)+a^5K_1(a)\right.\nonumber\\
&&+2b\left(3a^2K_2(a)+a^3K_1(a)-3a^3K_3(a)+a^4K_2(a)\right)\nonumber\\
&&\left.b^2\left(3a^2K_2(a)+a^3K_1(a)\right)\right]
\eea

\begin{figure}[ht]
\includegraphics[width=\columnwidth, height = 7.0cm]{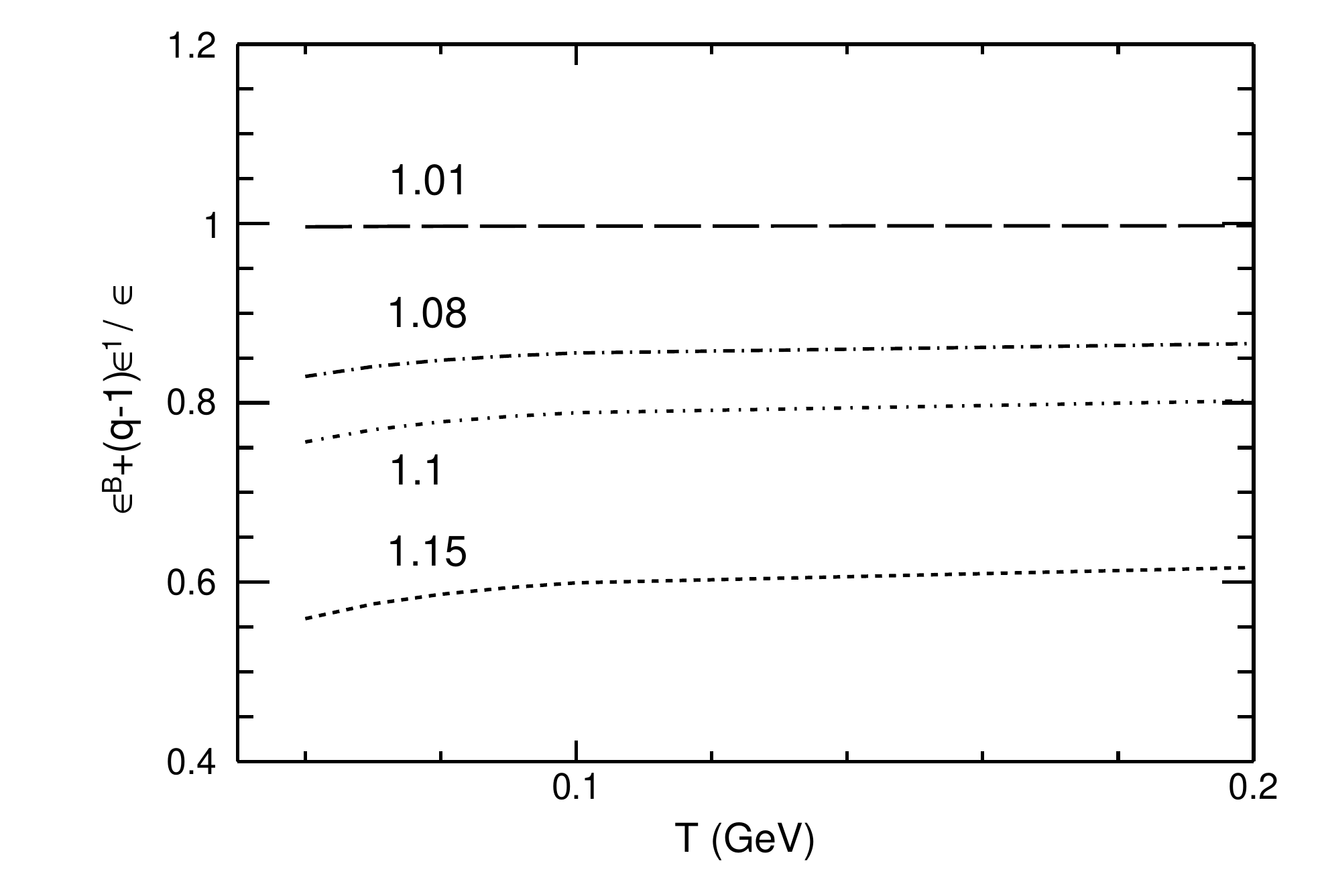}		
\caption{\label{ene_density_tsallis_caption} 
The ratio of the energy density calculated to first order in $(q-1)$ normalized to the energy density of a Tsallis gas as
a function of the temperature for different values of the parameter $q$.  The mass is taken as being the pion mass. 
The values of the parameter $q$ are 1.01
for the dashed line, 1.08 for the dot-(long)dashed line, 1.1 for the
dot-(short)dashed line and 1.15 for the dotted line.
}
\label{fig:energy_density_tsallis}
\end{figure}

\begin{figure}[ht]
\includegraphics[width=\columnwidth, height = 7.0cm]{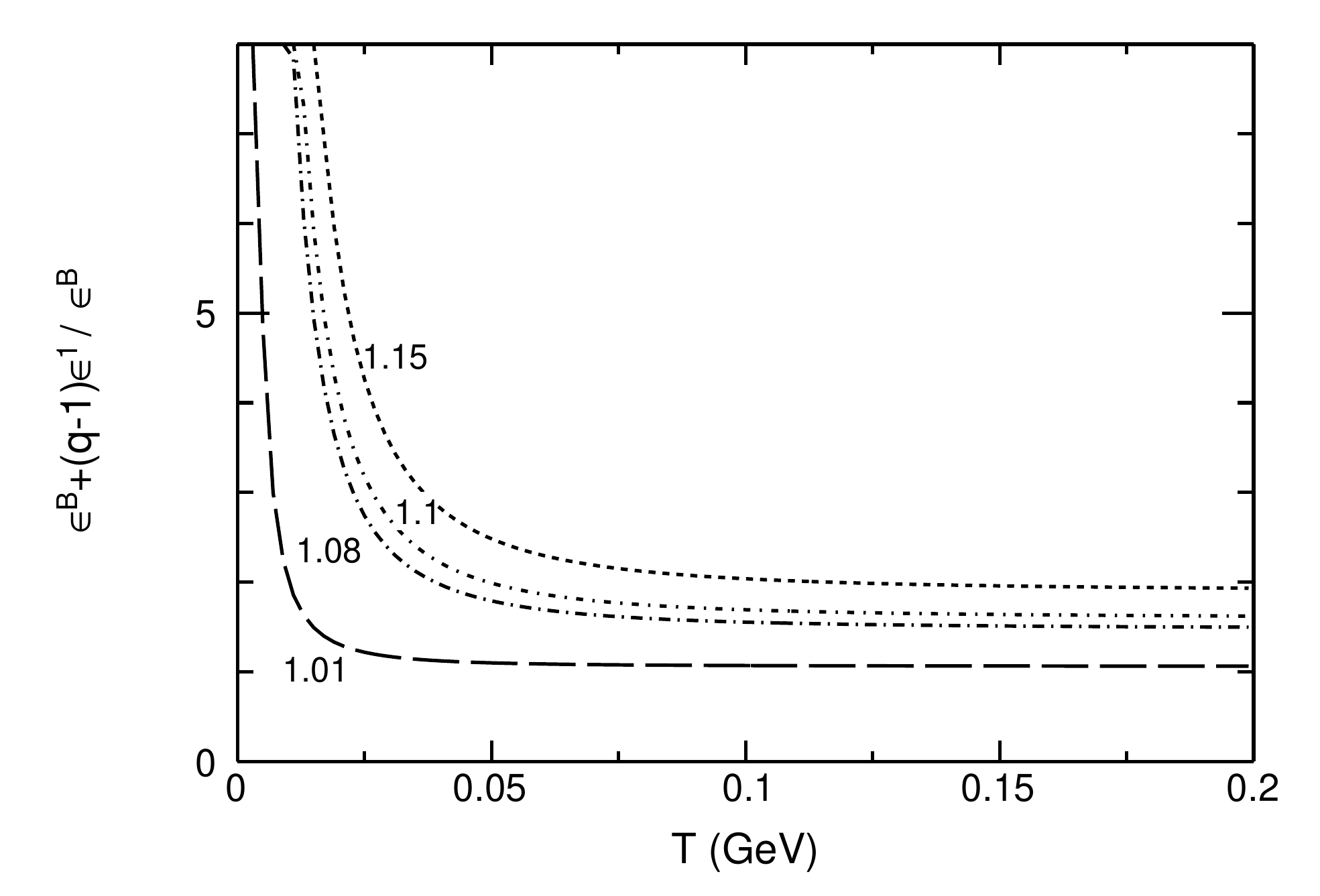}		
\caption{\label{energy_density_boltzmann_caption} 
The ratio of the energy density calculated to first order in $(q-1)$ normalized to the energy density of a Boltzmann gas as
a function of the temperature for different values of the parameter $q$.
}
\label{fig:energy_density_boltzmann}
\end{figure}

In Fig.~\ref{fig:energy_density_tsallis} we show the ratio of the energy density to first order in $(q-1)$ 
to the full energy density as given by the Tsallis distribution, $(\epsilon^B+(q-1) \epsilon^1)/\epsilon$ for several values of $q$  indicated in the   figure
as a function of the temperature $T$.\\
Again, as noted previously for the particle density, it can be seen that the expansion in $(q-1)$ is excellent 
if $(q-1) = 0.01$ but rapidly deviates from the full Tsallis distribution for larger values of $q$.
Also here, for $(q-1) \approx 0.1$ the deviations are of the order of 20 \% as can be seen from Fig.~\ref{fig:energy_density_tsallis}.
%
%
For comparison we show in Fig. \ref{fig:energy_density_boltzmann}, the first order expansion compared to the Boltzmann expression,   $(\epsilon^B+(q-1)n^1)/\epsilon^B$, 
as a function of the temperature $T$ for several values of the parameter $q$. As in the previous case the deviations are most pronounced for small values of the temperature. 
%
Finally, the pressure is given by
\begin{equation}
 P^B + (q-1)P^1
\end{equation}

\begin{equation}
P^B = \frac{g\mathrm{e}^{\frac{\mu}{T}}T^4 a^2 K_2(a)}{2\pi^2}
\end{equation}

\bea
P^1&=& \frac{g\mathrm{e}^{\frac{\mu}{T}}T^4}{4\pi^2} \left[a^4K_2(a)+3a^3K_3(a)\right. \nonumber\\ 
&&\left. -2a^3bK_3(a) +a^2b^2K_2(a)+2a^2bK_2(a)\right]
\eea
 
\begin{figure}[ht]
\includegraphics[width=\columnwidth, height = 7.0cm]{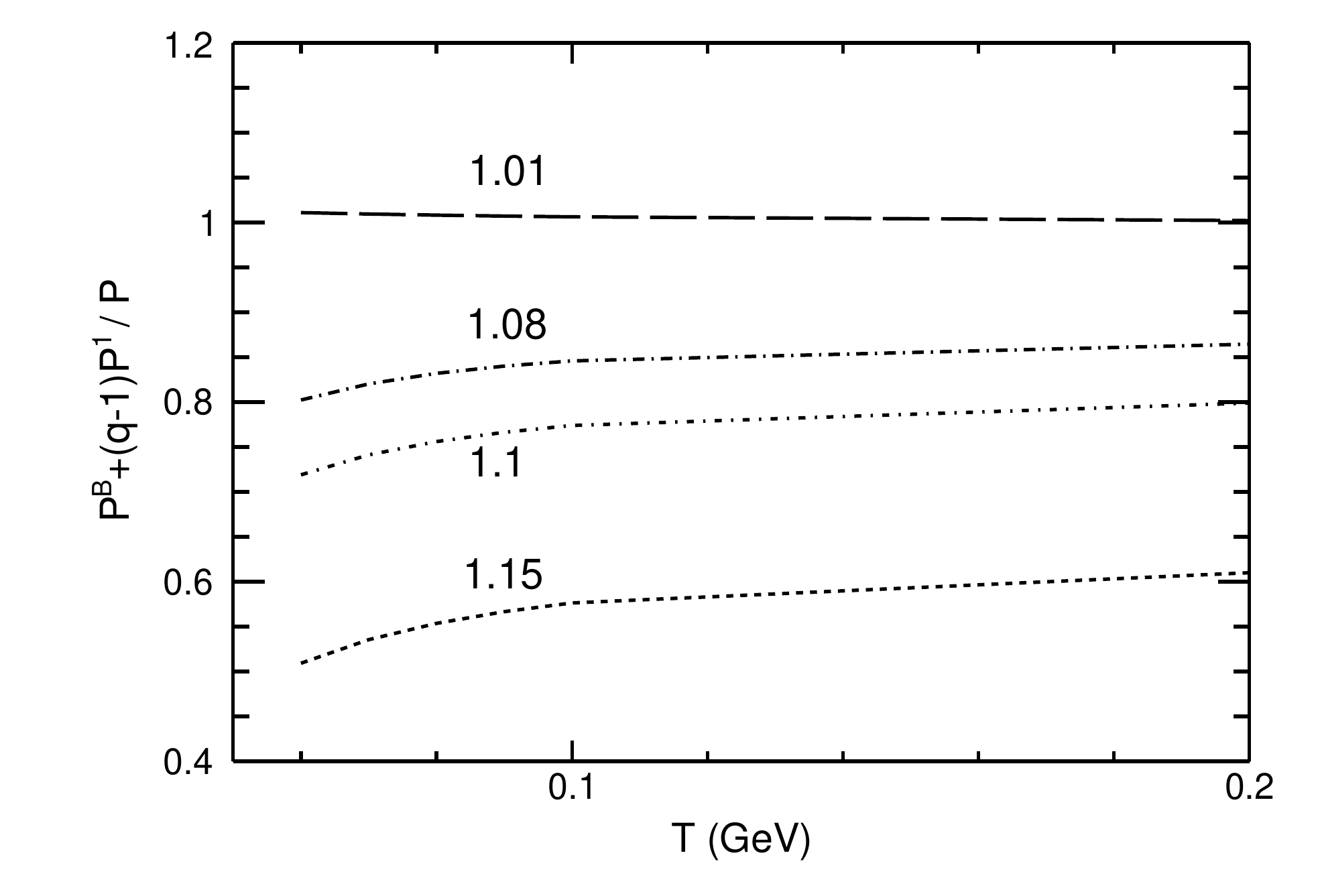}		
\caption{\label{fig:pressure_tsallis_caption} 
The ratio of the pressure calculated to first order in $(q-1)$ normalized to the pressure of a Tsallis gas as
a function of the temperature for different values of the parameter $q$.
}
\label{fig:pressure_tsallis}
\end{figure}

\begin{figure}[ht]
\includegraphics[width=\columnwidth, height = 8.0cm]{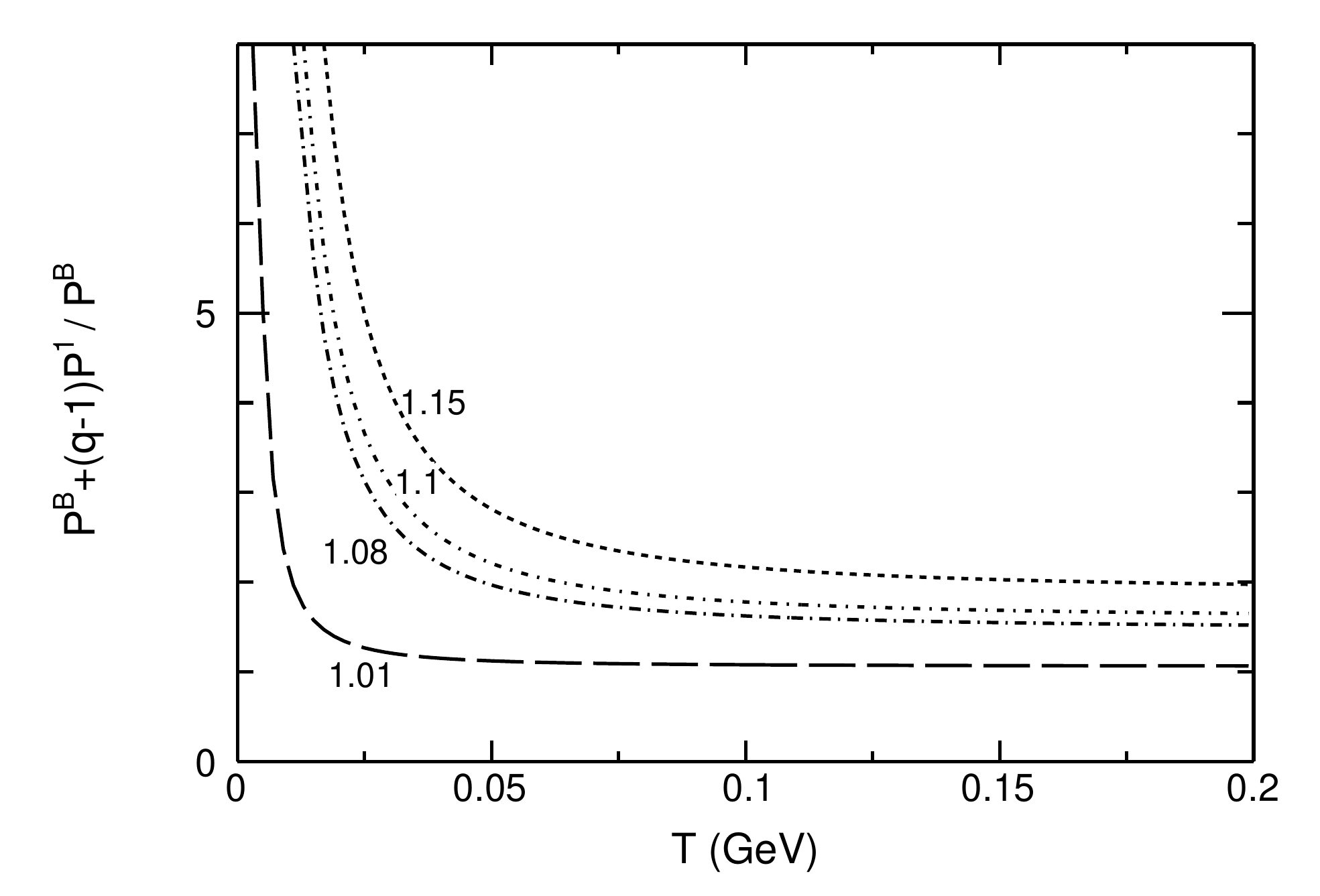}		
\caption{\label{fig:pressure_caption_boltzmann} 
The ratio of the pressure calculated to first order in $(q-1)$ normalized to the pressure of a Boltzmann gas as
a function of the temperature for different values of the parameter $q$.
\label{fig:pressure_boltzmann}
}
\end{figure}

In Fig.~\ref{fig:pressure_tsallis} we show the ratio of the pressure to first order in $(q-1)$ 
to the full pressure as given by the Tsallis distribution, $(P^B+(q-1)P^1)/P$ for several values of $q$  indicated in the   figure
as a function of the temperature $T$.\\
Again, as noted previously, it can be seen that the expansion in $(q-1)$ is excellent 
if $(q-1) = 0.01$ but rapidly deviates from the full Tsallis distribution for larger values of $q$.
Also here, for $(q-1) \approx 0.1$ the deviations are of the order of 20 \% as can be seen from Fig.~\ref{fig:pressure_tsallis}.
%
%
%
%
%
Again, we show in Fig. \ref{fig:pressure_boltzmann}, the first order expansion compared to the Boltzmann expression,   $(P^B+(q-1)P^1)/P^B$, as a function of the temperature 
$T$ for several values of the parameter $q$. In this case the deviations are most pronounced for small values of the temperature. 

The entropy density can be obtained from the above expressions by using the thermodynamic relation
\begin{equation}
\epsilon + P = T s +\mu n
\end{equation}
and is not being shown here.

Before closing the section, we comment about the validity of the present
expansion up to the second order in $(q-1)$.  As we have already seen, for truncation of the expansion up to the
first order with pure Tsallis distribution, one needs to satisfy two
conditions,  i.e.
\begin{equation}
|1-q|\frac{E}{T}<1
\label{con1}
\end{equation}
and
\begin{equation}
|1-q|\left(\frac{E}{T}\right)^2<2
\label{con2}
\end{equation}

With the thermodynamically consistent Tsallis distribution, 
the second condition turns out to be $|q(1-q)|(E/T)^2\\<2$. For expansion up
to second order in $(q-1)$ with modified Tsallis distribution, the condition
becomes $q^2|1-q|(E/T)^3<3$.

Given the two values (the highest and the lowest) of $(q-1)$ used in the present
analysis, we want to put an upper bound in $E/T$ until which the expansion will
be reliable. If $(q-1)=0.15$, Eq.\eqref{con1} gives $E/T<6.7$ and the modified
condition for expansion up to first order gives $(E/T)^2<11.59$. That means for
$T=0.1$ GeV, the expansion will be reliable up to $E\approx0.3$ GeV when
$(q-1)=0.15$.

\begin{figure}[ht]
\includegraphics[width=\columnwidth, height = 8.0cm]{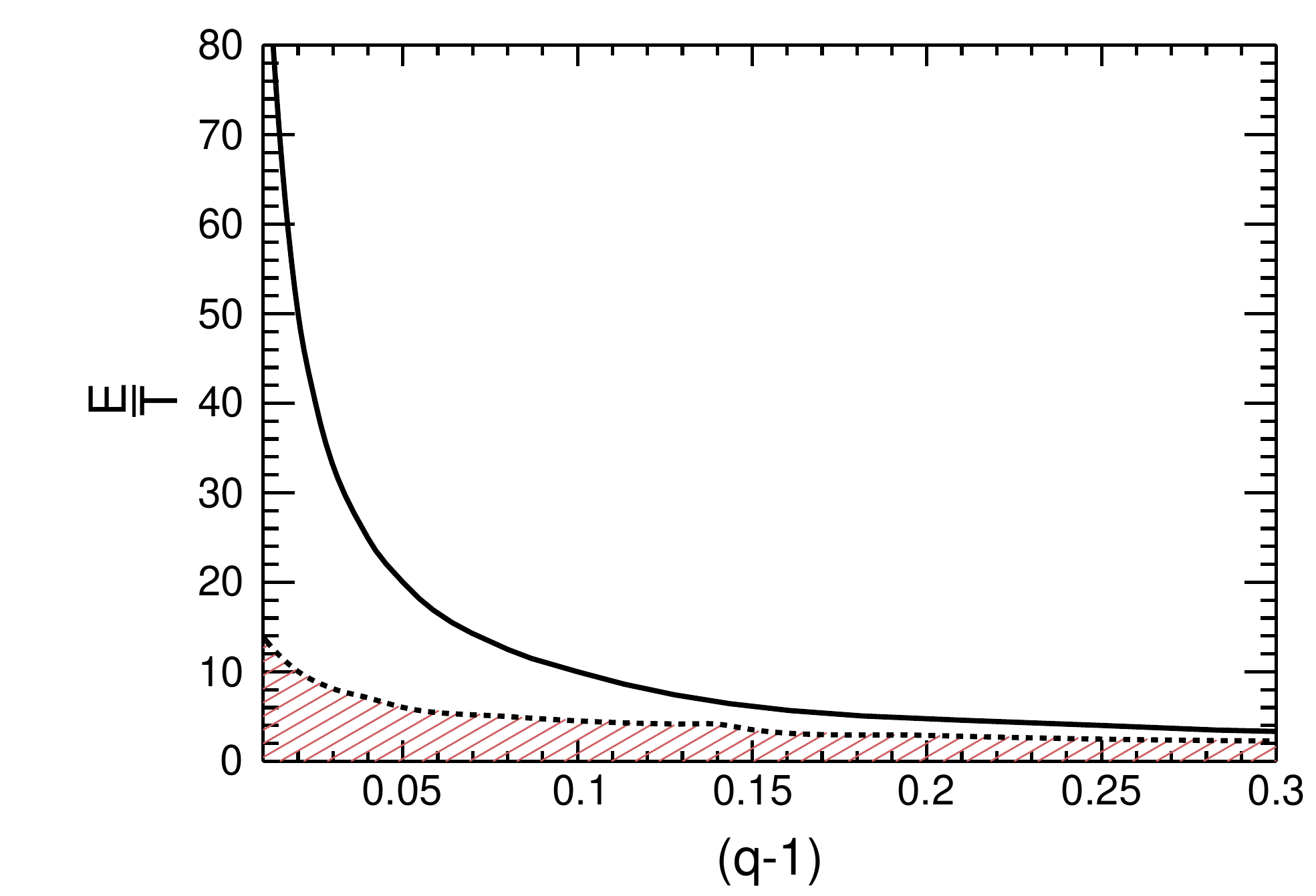}
\caption{(colour online) The region of validity for the expansion in $(q-1)$ up to second order. The area 
under the solid line denotes the region where only the condition
$|1-q|E/T<1$ is satisfied. The common overlapped area (i.e. the area under the dotted line)
depicts the region within which all the conditions for expansion up to second order
are satisfied.
\label{regofval}}
\end{figure}

If $(q-1)=0.01$, Eq.\eqref{con1} gives $E/T<100$ and the modified condition for
expansion up to first order gives $(E/T)^2<198$. That means for $T=0.1$ GeV, the
expansion will be reliable up to $E\approx1.4$ GeV when $(q-1)=0.01$. The
permissible values of $(q-1)$ and $E/T$ for reliable expansion up to the second
order in $(q-1)$ are shown to reside inside the smaller area filled with slanted
lines. This is shown in Fig. \ref{regofval}.

Hence, we conclude from the above discussion that the smaller the $q$ value
the more reliable the expansion becomes.

\section{Inclusion of Flow to First Order in $(q-1)$}
In order to see how the inclusion of flow could improve the description of the
transverse momentum distributions obtained in Pb-Pb collisions, we have included
a constant flow velocity, $\beta$. Assuming space-like freeze-out surface, the
invariant yield is given by (see Appendix for the derivation). 
\bea
\frac{1}{p_T}\frac{dN}{dp_Tdy} &&= \frac{gV}{(2\pi)^2} \nn\\
&&\biggl\{ 2 T [ r I_0(s) K_1(r) - s I_1(s) K_0(r) ]\nn\\
&&-(q-1) T r^2  I_0(s) [K_0(r)+K_2(r)] \nn\\
&&+ 4(q-1)~T rs I_1(s) K_1(r)\nn\\
&&-(q-1)Ts^2 K_0(r)[I_0(s)+I_2(s)]\nn\\
&& + \frac{(q-1)}{4}T r^3 I_0(s) [K_3(r)+3K_1(r)]\nn\\
&&-\frac{3(q-1)}{2}T r^2 s [K_2(r)+K_0(r)] I_1(s) \nn\\
&&+ \frac{3(q-1)}{2} T s^2 r [I_0(s)+I_2(s)] K_1(r) \nn\\
&&-\left.\frac{(q-1)}{4}T s^3 [I_3(s)+3I_1(s)] K_0(r)\right\}
\label{tsallisflow}
\eea
where 

\bea
r\equiv\frac{\gamma m_T}{T} \\
s\equiv\frac{\gamma v p_T}{T}
\eea
$I_n(s)$ and $K_n(r)$ are  the modified Bessel functions of the first
and second kind. Now, in this formula, the freeze-out surface has been
considered to be space-like and so the integration over the freeze-out surface
turns out to be trivial. For a more detailed treatment of the freeze-out
surface in this context, the readers are referred to Ref. \cite{urmossy}.

\begin{figure}[ht]
\includegraphics[width=0.9\columnwidth, height = 14.0cm]{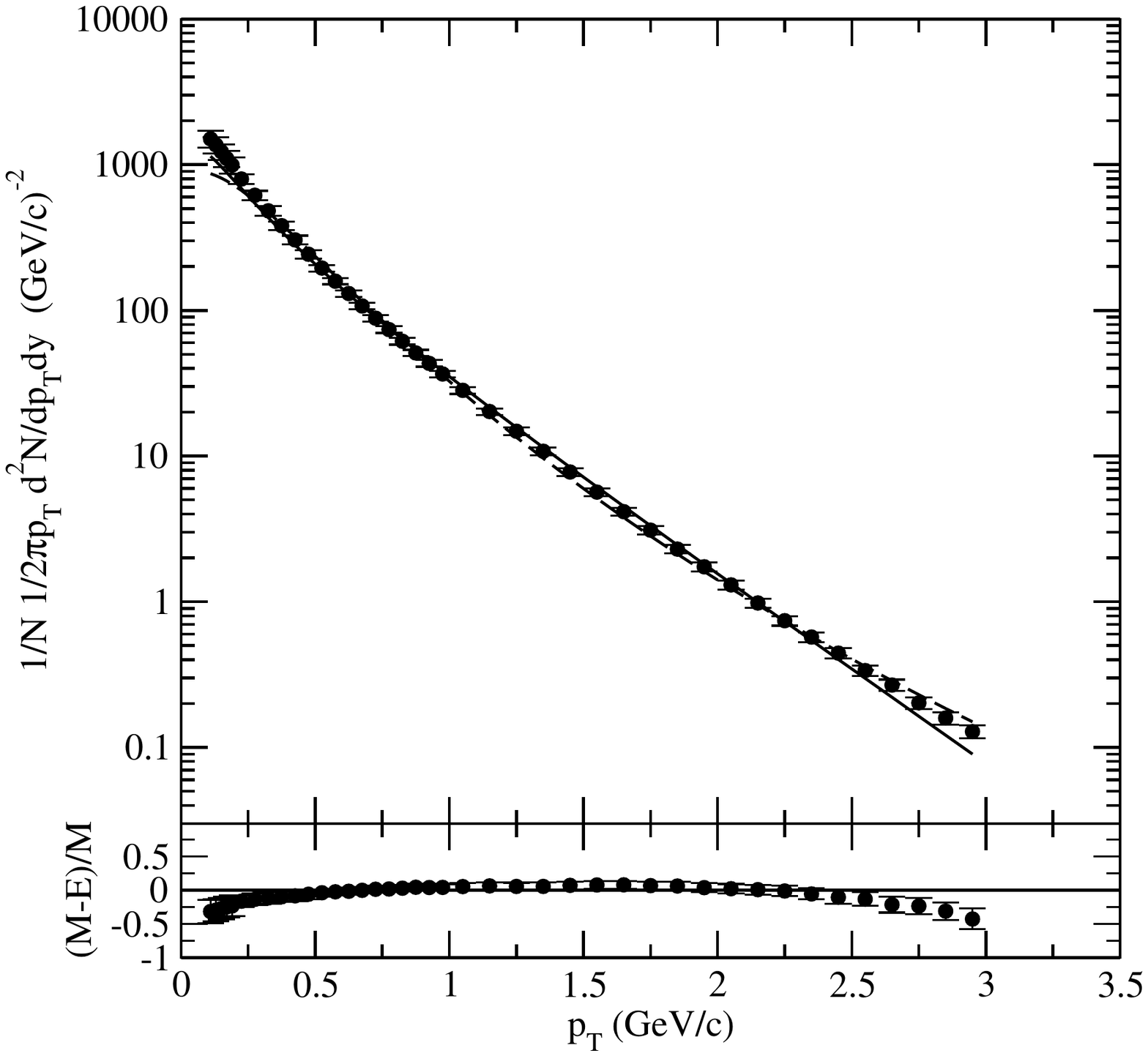}		
\caption{\label{fig:ALICEpp} Fits to the normalized differential $\pi^-$
yields as measured by the ALICE collaboration in  $(0-5)\%$ Pb-Pb collisions at
$\sqrt{s_{\rm NN}}$ = 2.76 TeV~\cite{ALICE_PbPb}. The fit with the Tsallis
distribution including flow  keeping terms to first  order in $(q-1)$
(dashed line). The flow velocity is fixed at $\beta = 0.609$, with $T = 146 $ MeV, $q = 1.030$ and the 
radius of the volume is $R = 29.8$ fm. The solid line is the Tsallis
distribution without flow as given in Fig. ~\ref{fig:pbpb}. The lower part of
the figure shows the difference between model (M), i.e. Tsallis with flow up to
first order in $(q-1)$, and experiment (E) normalized to the model (M) values.
}
\label{fig:flow}
\end{figure}
The comparison between model and experiment is quite good with notable
deviations at small values of the transverse momentum $p_T$ and again above
values of 2.5 GeV (see Fig. \ref{fig:flow}).  These could easily be attributed to
the coarse way of treating transverse flow. More detailed investigations have
been carried out in~\cite{TBW}.

\section{Summary and Conclusion}
\label{summary}
The Tsallis distribution describes extremely well the transverse momentum
distributions in p - p collisions at high energies. All fits performed so far
show that the parameter $q$ is always close to 1. In view of this, we have
presented in this paper a series expansion of quantities relevant in the
analysis of high energy physics in $(q-1)$. The Tsallis distribution itself has
been obtained to second order in $(q-1)$. A rough comparison with experimental
data has been done for the transverse momentum distributions obtained in p-p
collisions. In the case of Pb-Pb collisions we have given an estimate using flow
with a fixed flow velocity. In most cases the series expansion turns out not to
be a useful description of transverse momentum distributions but it could be
useful in analyses where a comparison and detailed investigation is needed when
comparing the Tsallis and the Boltzmann distributions. Furthermore, a
systematic study of the identified particle yield due to $p-p$ and heavy-ion
collision at RHIC and LHC has been done in \cite{dhananjay} where the flow
formula (Eq. \eqref{tsallisflow}) has been used to fit the spectra.

\section*{Acknowledgement} The authors would like to thank Dr. Prakhar
Garg, IIT Indore for useful discussions and help during the
preparation of the manuscript. We thank the referees for their helpful
comments which improved the paper.

\appendix


\section*{Appendix}
 \section{Derivation of Eq. \eqref{dndy} and Eq. \eqref{fittingformula}}
 \label{eq1314}

The result quoted in Eq. \eqref{dndy} can be obtained
by a change of  variable from $p_T$ to $x$ which is defined by
\begin{equation}
 x \equiv 1 + (q-1)\frac{m_T-\mu}{T} .
 \end{equation}
 This  leads to the following integral
 \begin{eqnarray}
 \frac{dN}{dy}\bigg|_{y=0} &=&  \int_0^\infty
dp_T~\frac{dN}{dp_Tdy}\bigg|_{y=0} \nonumber\\
                           &=&\frac{gV}{(2\pi)^2}\int_0^\infty
dp_T~p_Tm_T\left[1 + (q-1)\frac{m_T-\mu}{T}\right]^{-\frac{q}{q-1}} \nonumber\\
                           &=&\frac{gV}{(2\pi)^2}\frac{T}{q-1}\int_a^\infty dx~
x^{-\frac{q}{q-1}} \left[\frac{x-1}{q-1}T + \mu\right]^2 \nonumber\\
 \end{eqnarray}
 with
 \begin{equation}
 a \equiv 1 + (q-1)\frac{m-\mu}{T}
 \end{equation}
 
 This is an elementary integral over a polynomial function. The result is

 \begin{eqnarray}
 \frac{dN}{dy}\bigg|_{y=0} &=&\frac{gV}{(2\pi)^2}\left[1 +
(q-1)\frac{m-\mu}{T}\right]^{-\frac{1}{q-1}} \nonumber\\
                           &&\left\{\frac{T^3}{(2q-3)(q-2)}\left[2 -
(q-2)\left(\frac{m-\mu}{T}\right)^2 + 2\frac{m-\mu}{T}\right]\right. \nonumber\\
&&\left.-2\frac{T^2\mu}{q-2}\left[1+\frac{m-\mu}{T}\right] +T\mu^2\right\}
\label{dndyapp}
 \end{eqnarray}
This makes it possible to express the volume $V$ in terms of
$\frac{dN}{dy}\bigg|_{y=0}$ from Eq. \eqref{dndy} in the text.
Replacing the volume term in Eq. \eqref{tsallisptdist} in terms of
Eq. \eqref{dndyapp}, we get Eq. \eqref{fittingformula}.

\section{Expansion in ($q-1$) of Tsallis distribution}

\label{apptaylorsexp}
The Taylor expansion is done efficiently by the following change of variables:
\bea
q-1 &\equiv& x; \frac{E-\mu}{T} \equiv \Phi \\
1+(q-1)\frac{E-\mu}{T} &\equiv & 1+x ~\Phi \equiv f(x) \\
-\frac{q}{q-1} &=& -\frac{1+x}{x} \equiv~g(x)
\eea

\noi The relevant distribution function can then be written as

\bea
\left[1+(q-1)\frac{E-\mu}{T}\right]^{-\frac{q}{q-1}}  
&=& \left[1+x\Phi\right]^{-\frac{1+x}{x}} \nn\\ 
&=& f(x)^{g(x)} \nn\\
&=& \mathcal{F}(x)
\eea
Expanding $\mathcal{F}(x)$ in Taylor series about $x=0$, we get
\begin{eqnarray}
\mathcal{F}(x=0) &=& \mathrm{e}^{-\Phi} \\
\left.\frac{d {\mathcal{F}(x)}}{dx}\right|_{x = 0 } 
&=& \frac{1}{2} (-2+\Phi) \Phi\mathrm{e}^{-\Phi} \\
\frac{d^2 {\mathcal{F}(x)} }{dx^2}_{x = 0} &=& \frac{\Phi^2}{12}\left(24-20\Phi+3\Phi^2\right)\mathrm{e}^{-\Phi}
\end{eqnarray}
for $\mathcal{O}(x^0)$, $\mathcal{O}(x)$ and $\mathcal{O}(x^2)$ respectively. 
The final results can be summarized in the following equation
\begin{eqnarray}
&&\left[1+(q-1)\frac{E-\mu}{T}\right]^{-\frac{q}{q-1}}\nonumber\\
&\simeq& \mathrm{e}^{-\frac{E-\mu}{T}}\nonumber\\
&+& (q-1)\frac{1}{2}\frac{E-\mu}{T}\left( -2 + \frac{E-\mu}{T}\right)\mathrm{e}^{-\frac{E-\mu}{T}}\nonumber\\
&+& \frac{(q-1)^2}{2!}\frac{1}{12}\left[\frac{E-\mu}{T}\right]^2\nonumber\\
&\times& \left[ 24 - 20\frac{E-\mu}{T} +3\left( \frac{E-\mu}{T}\right)^2\right]\mathrm{e}^{-\frac{E-\mu}{T}} \nonumber\\
+ ....\\
\end{eqnarray}
For completeness we also quote the following result 
\begin{eqnarray}
&&\left[1+(q-1)\frac{E-\mu}{T}\right]^{-\frac{1}{q-1}}\nonumber\\
&\simeq& \mathrm{e}^{-\frac{E-\mu}{T}}\nonumber\\
&+& (q-1)\frac{1}{2}\left( \frac{E-\mu}{T}\right)^{2}\mathrm{e}^{-\frac{E-\mu}{T}}\nonumber\\
&+& \frac{(q-1)^2}{2!}\frac{1}{12}\left[\frac{E-\mu}{T}\right]^2\nonumber\\
&\times& \left[ -8\left(\frac{E-\mu}{T}\right) +3\left( \frac{E-\mu}{T}\right)^2\right]\mathrm{e}^{-\frac{E-\mu}{T}} \nn\\
+ ....
\end{eqnarray}

%

\section{Tsallis Thermodynamics}
\subsection{Particle number density $n$}

\label{appthermo}
Up to $\mathcal{O}(q-1)$, the number density can be written as:

\bea
n&=&\frac{g}{(2\pi)^3}\int d^3p \left[\mathrm{e}^{-\frac{E-\mu}{T}}+(q-1) 
\frac{E-\mu}{2T} \right.\nonumber\\ && \left(-2+\frac{E-\mu}{T}\right)
\left.\mathrm{e}^{-\frac{E-\mu}{T}}\right]
\eea

\noi We define, 

\bea
p^2+m^2=E^2 ;  \nonumber\\
\frac{E}{T}=\omega, \frac{m}{T}=a, \frac{\mu}{T}=b
\label{defvar}
\eea

\noi And hence,
\bea
n &=& n^B +\frac{g(q-1)\mathrm{e}^{\frac{\mu}{T}}T^3}{4\pi^2}\int d\omega~\omega \left(\omega^2-a^2\right)^\frac{1}{2}\nonumber\\
&&\left(-2\omega+2b+\omega^2+b^2-2\omega b\right)\mathrm{e}^{-\omega}\nonumber\\
n&=&n^B +\frac{g(q-1)\mathrm{e}^{\frac{\mu}{T}}T^3}{4\pi^2}\left[-6a^2K_2(a)-2a^3 K_1(a)\right.\nonumber\\
&& -4a^2bK_2(a)+3a^3K_3(a)+a^4K_2(a)+a^2 b^2 K_2(a)\nonumber\\
&&\left.-2a^3bK_1(a)\right]
\label{n}
\end{eqnarray}
\vspace{0.2in}

\noi Here the following form of the modified Bessel function of second kind \cite{abstegun} is used:

\bea
K_n(a)=\frac{2^{n-1}(n-1)!}{(2n-2)!a^n} \int_{a}^{\infty} d\omega 
~\omega (\omega^2-a^2)^{n-\frac{3}{2}} \mathrm{e}^{-\omega} 
\label{bessel1}
\eea
On the other hand, by using Boltzmann Statistics
one obtains,

\bea
n^B=\frac{g\mathrm{e}^{\frac{\mu}{T}}T^3 a^2 K_2(a)}{2\pi^2}
\eea


\subsection{Pressure}
To first order in ($q-1$) the pressure is given by:
\begin{eqnarray}
P=\frac{g}{(2\pi)^3}\int d^3p~\frac{p^2}{3E} \left[\mathrm{e}^{-\frac{E-\mu}{T}}+(q-1) \frac{E-\mu}{2T}\right.\nonumber\\
\left.\left(-2+\frac{E-\mu}{T}\right)
\mathrm{e}^{-\frac{E-\mu}{T}}\right]
\end{eqnarray}

\noi Using the same definitions as in Eq. \eqref{defvar}, we get

\bea
P &=& P^B+\frac{g(q-1)\mathrm{e}^{\frac{\mu}{T}}T^4}{12\pi^2}\int d\omega~ \left(\omega^2-a^2\right)^\frac{3}{2}\nonumber\\
&&\left(-2\omega+2b+\omega^2+b^2-2\omega b\right)\mathrm{e}^{-\omega}\nonumber\\
&=& P^B+\frac{g(q-1)\mathrm{e}^{\frac{\mu}{T}}T^4}{4\pi^2} \left[a^4K_2(a)+3a^3K_3(a)\right. \nonumber\\ 
&&\left. -2a^3bK_3(a) +a^2b^2K_2(a)+2a^2bK_2(a)\right]
\eea
Using Eq. \eqref{defvar} along with Eq. \eqref{bessel1} as well as another representation of the modified Bessel function,

\bea
K_n(a)=\frac{2^{n} n!}{(2n)! a^n} \int_{a}^{\infty} d\omega (\omega^2-a^2)^{n-\frac{1}{2}} \mathrm{e}^{-\omega} 
\label{bessel2}
\eea
Eq. \eqref{bessel1} can be obtained from Eq. \eqref{bessel2} by dint of partial integration.
The Boltzmann pressure density is, similarly given by: 

\bea
P^B = \frac{g\mathrm{e}^{\frac{\mu}{T}}T^4 a^2 K_2(a)}{2\pi^2}
\eea

\subsection{Energy density $\epsilon$}

\bea
 \epsilon &=& \frac{g}{(2\pi)^3}\int d^3p~E\left[\mathrm{e}^{-\frac{E-\mu}{T}}+(q-1) \frac{E-\mu}{2T} \right.\nonumber\\  
 && \left(-2+\frac{E-\mu}{T}\right)  \left.\mathrm{e}^{-\frac{E-\mu}{T}}\right]
\eea
Using Eq. \eqref{defvar} along with Eq. \eqref{bessel1},

\bea
\epsilon 
&=&\epsilon^B + \frac{g(q-1)\mathrm{e}^{\frac{\mu}{T}}T^4}{4\pi^2}\left[9a^3K_3(a)+4a^4K_2(a)+a^5K_1(a)\right.\nonumber\\
&&+2b\left(3a^2K_2(a)+a^3K_1(a)-3a^3K_3(a)+a^4K_2(a)\right)\nonumber\\
&&\left.b^2\left(3a^2K_2(a)+a^3K_1(a)\right)\right]
\eea
with 

\bea
\epsilon^B =  \frac{ g \mathrm{e}^{\frac{\mu}{T}} T^4}{2\pi^2} (3a^2K_2(a)+a^3K_1(a))
\eea

\subsection{Checking Thermodynamic Consistency}

We know that for thermodynamic variables $n$ and $P$ at non-zero $\mu$ we have,
\bea
n &=& \partial{P}/\partial{\mu} \label{np}\\
n^B+(q-1)n^1+\ldots  &=& \partial{P^B}/\partial{\mu} + \nn\\
&& (q-1)\partial P^1/\partial\mu + \ldots
\eea
where the superscripts denote the order of ($q-1$) in the expansion. Since the terms in the expansion are
linearly independent, the identity given by Eq. \eqref{np} is to be satisfied at every order of ($q-1$). Since
our expansion is up to $\mathcal{O}(q-1)$, we will check the identity for that order now. Rearranging and reordering 
the $\mathcal{O}(q-1)$ terms in Eq. \eqref{n} we get,

\bea
n^1 &=&  \frac{g\mathrm{e}^{\frac{\mu}{T}}T^3}{4\pi^2} \left[  a^3K_3(a)+a^4K_2(a)\right.\nonumber\\
&& -4a^2bK_2(a) +2a^3b ( K_3(a)-K_1(a) ) \nonumber\\ 
&& -2a^3bK_3(a)+a^2b^2K_2(a)-6a^2K_2(a) \nn\\ 
&+& \left.  2a^3 ( K_3(a)-K_1(a) ) \right]\nonumber\\
\eea
Using the recursion relation for the modified Bessel's functions,

\bea
K_{n+1}(a)-K_{n-1}(a)=\frac{2n}{a}K_n(a)
\eea
 
\bea 
n^1 &=&  \frac{g(q-1)\mathrm{e}^{\frac{\mu}{T}}T^3}{4\pi^2} \left[ a^3K_3(a)+a^4K_2(a)\right.\nonumber\\ 
&& \left. +4a^2bK_2(a)-2a^3bK_3(a)+a^2b^2K_2(a) \right.\nn\\
&& \left. +2a^2 K_2(a) \right] \nn\\
&=& \frac{\partial P^1}{\partial \mu}
\eea
Hence, proved.

\section{Momentum Distribution}
The invariant particle yield is given by,
\begin{eqnarray}
E \frac{dN}{d^3p}=~\mathcal{C} E\left[1+(q-1)\frac{E-\mu}{T}\right]^{-\frac{q}{q-1}}
\end{eqnarray}
Where $\mathcal{C}=\frac{gV}{(2\pi)^3}$. Assuming $q-1\ll1$ we are to expand it in Taylor series. Let, 
\begin{eqnarray}
\frac{dN}{p_Tdp_Tdyd\phi}&=&\mathcal{C}E\left[\mathrm{e}^{-\Phi}+\frac{x}{1!}\frac{1}{2}
\left(-2+\Phi\right)\Phi\mathrm{e}^{-\Phi}\right.\nonumber\\
&&\left.+\frac{x^2}{2!}\frac{\Phi^2}{12}\left(24-20\Phi+3\Phi^2\right)\mathrm{e}^{-\Phi}+\cdots\right]\nonumber\\
\end{eqnarray}
where $y$ is rapidity and $\phi$ is the azimuthal angle of emission. Now, parameterizing energy $E$ in terms of 
rapidity $y$  i.e. putting $E=m_T\mathrm{cosh}y$ and integrating over $y$ we get Eq. \eqref{trmassspec}.

\section{Flow in Tsallis Distribution}
\label{app:flow}
We use the following ansatz (in cylindrical polar coordinates) for introducing flow inside our calculations:

\begin{eqnarray}
p^{\mu} &=& (m_T\mathrm{cosh}y, p_T\mathrm{cos}\phi, p_T\mathrm{sin}\phi, m_T\mathrm{sinh}y) \\
u^{\mu} &=& (\gamma \mathrm{cosh}\zeta, \gamma v \mathrm{cos}\alpha, \gamma v \mathrm{sin}\alpha, 
\gamma \mathrm{sinh}\zeta)
\end{eqnarray}
where $(\zeta)y$ is the (space-time)rapidity of particles (fluid-element) and $v$ is the velocity of fluid. Now, to include
flow inside the Tsallis distribution, we replace $E\rightarrow p^{\mu}u_{\mu}$ assuming temperature to be
scalar. The dot product then becomes,
\be
p^{\mu}u_{\mu}=\gamma m_T \mathrm{cosh}(y-\zeta)-\gamma v p_T \mathrm{cos}(\phi-\alpha)
\label{pdotu}
\ee

Now putting Eq. \eqref{pdotu} in Eq. \eqref{order12} up to $\mathcal{O}(q-1)$ and integrating over 
$\phi$ and $\zeta$ we get,
\bea
\frac{1}{p_T}\frac{dN}{dp_Tdy} &&= \frac{gV}{(2\pi)^2} \nn\\
&&\left\{ 2 T [ r I_0(s) K_1(r) - s I_1(s) K_0(r) ]\right.\nn\\
&&-(q-1) T r^2  I_0(s) [K_0(r)+K_2(r)] \nn\\
&&+ 4(q-1)~T rs I_1(s) K_1(r)\nn\\
&&-(q-1)Ts^2 K_0(r)[I_0(s)+I_2(s)]\nn\\
&& + \frac{(q-1)}{4}T r^3 I_0(s) [K_3(r)+3K_1(r)]\nn\\
&&-\frac{3(q-1)}{2}T r^2 s [K_2(r)+K_0(r)] I_1(s) \nn\\
&&+ \frac{3(q-1)}{2} T s^2 r [I_0(s)+I_2(s)] K_1(r) \nn\\
&&-\left.\frac{(q-1)}{4}T s^3 [I_3(s)+3I_1(s)] K_0(r)\right\}
\eea
where 

\bea
r\equiv\frac{\gamma m_T}{T} \\
s\equiv\frac{\gamma v p_T}{T}
\eea
$I_n(s)$ is the modified Bessel function of first kind.


\end{document}